\documentstyle[11pt,twoside]{article}
\begin{document}\hbadness=10000\thispagestyle{empty}
\pagestyle{myheadings}
\markboth{H.-Th. Elze}
{Relativistic Quantum Transport Theory}
\title{{\bf Relativistic Quantum Transport Theory}}
\author{$\ $\\
{\bf Hans-Thomas Elze}\\ $\ $\\
Universidade Federal do Rio de Janeiro, Instituto de F\'{\i}sica\\
\,Caixa Postal 68.528, 21945-970 Rio de Janeiro, RJ, Brazil}
\vskip 0.5cm
\date{March 2002}
\maketitle
\vspace{-8.5cm}
\vspace*{8.0cm}
\begin{abstract}{\noindent
Relativistic quantum transport theory has begun to play an important 
role in the space-time description of matter under extreme conditions 
of high energy density in out-of-equilibrium situations. The following 
introductory lectures on some of its basic concepts and methods comprise 
the sections: 1. Introduction; 2. Aims of transport theory (classical); 
3. Quantum mechanical distribution functions - the density matrix 
and the Wigner function; 4. Transport theory for quantum fields; 
5. Particle production by classical fields; 6. Fluid dynamics of 
relativistic quantum dust.
\\
\noindent
}\end{abstract}

\section{Introduction}
The study of the behavior of matter under more and more extreme conditions has 
a long tradition motivated by the quest for understanding the forces among 
its constituents on smaller and smaller scales. Not only the attempts to understand 
such spectacular phenomena as the stellar supernova explosions or theories of even the 
primordial stages of cosmological evolution, but also the ever increasing collision energy 
of high-energy particle accelerators and the heavy-ion programs at CERN 
and RHIC, in particular, witness the most recent stages of this scientific development.  
  
In the latter context, often the complicated space-time dependence of what is really a 
quantum many-body system or what are highly dynamical interacting quantum fields is described  
in terms of a perfect fluid model. Since the seminal work by Fermi and Landau this
approach has been applied successfully, in order to study global features, such as multiplicity distributions
and apparently thermal transverse momentum spectra of produced particles, in high-energy collisions of strongly
interacting matter \cite{Fermi,Landau,Bj,QMs}. Similarly, the hydrodynamic approximation is often invoked in
astrophysical applications and cosmological studies of the early universe \cite{Weinberg}

The limitations of and likely necessary corrections to the fluid picture, however, have 
rarely been explored in the microscopic or high energy density domain. Difficulties reside 
in the derivation of consistent transport equations and in the amount of computational work required
to find realistic solutions; see Refs.\,\cite{EH,Cooper98,Blaizot99}, for example, for a review
and recent progress concerning selfinteracting scalar particles and the quark-gluon
plasma, respectively. More understanding of related hydrodynamic behavior, if any,
seems highly desirable.

For example, it has recently been shown that a free scalar field indeed behaves like a 
perfect fluid in the semiclassical (WKB) regime \cite{DomLev00}. More generally, the mechanisms of quantum
decoherence and thermalization in such systems which can be described hydrodynamically, i.e.
the emergence of classical deterministic evolution from an underlying quantum field theory,
are of fundamental interest \cite{WHZ90,I95,Lisewski99,BrunHartle99}.    
  
Having said this, it becomes obvious that not only particular applications of relativistic quantum 
transport theory motivated by experiments or observations -- such as a suitable quark-gluon plasma 
transport theory to be applied to the phenomenology of high-energy heavy-ion collisions, or a transport 
theory for the electro-weak interactions of the intense neutrino flux from a supernova core with its 
electron-positron plasma sphere -- are of interest, but that many interesting conceptual problems 
can be found in this field. It is the aim of the present introduction to describe some of its basic 
concepts and methods. 
   
It seems worth while to emphasize here that transport theory by its very nature aims to 
describe highly dynamical systems where the time dependence of the phenomena to be studied 
cannot be neglected. Therefore, one necessarily has to go beyond (thermal) equilibrium 
field theory, for example. A partial exception consists in linear response theory, where standard 
field theory methods are employed to calculate the short-time response of the system to 
necessarily small perturbations.   
 
The plan of these lectures is as follows. 
In Section 2 we present the motivation for transport theory by taking a cursory look at 
classical relativistic transport theory and its relation to relativistic hydrodynamics. 
In Section 3 we introduce quantum mechanical distribution functions. Especially, the need  
for the density matrix formalism is reviewed in basic terms, and the Wigner function is 
introduced. In Section 4, as an example, we develop the transport theory for the particular 
model of interacting quantum fields with a global O(4) symmetry, i.e. the linear sigma model, 
in the Hartree approximation. In Section 5 particle production by classical fields is described, presumably an 
important effect during early stages of heavy-ion collisions, solving fermion quantum transport 
equations perturbatively. Finally, in Section 6, we investigate the fluid dynamical behavior of 
relativistic quantum dust, solving the free quantum transport equations for arbitrary initial 
conditions exactly.

\section{Aims of transport theory}
Concerning the historical development as well as the systems under study, the 
subject matter of transport theory is most frequently associated with 
nonequilibrium plasmas of all sorts: 
\begin{itemize} 
\item The quark-gluon plasma with QCD interactions formed shortly after the Big Bang 
initiating the observable Universe and possibly recreated during high-energy  
collisions, in particular with heavy nuclear projectiles/targets ({\bf R+Q}). 
\item The $\nu\bar\nu e^+e^-$ plasma with electro-weak interactions created during supernova explosions between the proto-neutron star core and the leftover outer layers of the collapsing star ({\bf R}). 
\item $H$, $He,\;\dots\;$ fusion plasmas in burning stars. 
\item Electrodynamic plasma phenomena in the Earth's ionosphere leading to polar lights and  
thunderstorms with lightnings. 
\item Discharge plasmas used in neon lighting and plasma welding. 
\item The $e^-$ plasma in metals or semiconductors, the study of which has been advanced in solid state physics with particular attention to quantum effects ({\bf Q}).
\end{itemize} 
Here we marked by {\bf R} and/or {\bf Q} the systems where relativistic and/or quantum 
effects are known or expected to play an essential role. --
Abstracting from these examples, we notice two common qualitative features among these 
systems:
\begin{itemize}
\item microscopic distance scales (average interparticle distance $n^{-1/3}$, mean free 
path $\lambda_{mfp}$, etc.) \\
$\approx$ homogeneity scale $L$;
\item microscopic time scales (average lifetime $\tau_l$, relaxation time(s) $\tau_r$, etc.) \\ 
$\approx$ hydrodynamic time scale $\tau_h=L/c_s$, where $c_s$ denotes the sound velocity. 
\end{itemize}
Clearly, there will be exceptions to these qualitative statements and more 
precise characterizations of the plasma state can be given in the individual cases.  
However, this may suffice here, and we embark on the more formal description in the following. 
 -- We refer the reader to the monograph on relativistic kinetic theory of Ref.\,\cite{deGroot}, 
which presents an excellent detailed exposition of the more traditional material in this context. 
 
\subsection{Classical phase space description of many-body dynamics}    
For a classical many-particle system, consider the probability to find a 
particle in the 8-dimensional phase space volume element at the (four-vector) position $x$ 
with (four-vector) momentum $p$, 
\begin{equation}\label{dP}
\mbox{d}P(x,p)\equiv f(x,p)\mbox{d}^4x\mbox{d}^4p
\;\;, \end{equation} 
where $f(x,p)$ denotes the corresponding Lorentz scalar phase space density. Generally, in a classical system, 
the four-momentum is on-shell, such that not all components of $p$ are independent. In particular, 
we assume that the constraint expressing the energy  
in terms of the three-momentum and particle mass, $p^0=+\sqrt{\vec p^2+m^2}$, is incorporated in $f$. 
--  
Except when explicitly stated, our units are such that 
$\hbar =c=k_B=1$, and we use the Minkowski metric $g_{\mu\nu}=\mbox{diag}(1,-1,-1,-1)$. 
 
A remark is in order here. 
Clearly, in order to learn not only about the behavior of a typical particle but also 
about its correlations with others, one principally should study the \underline{one-body density} $f$ 
along with a two-body distribution $f(x_1,x_2,p_1,p_2)$, three-body distribution $\dots\;$, etc. 
This is quite complicated in general and we restrict our attention 
to the one-body density here. Sometimes it is useful to visualize this  
function alternatively as describing a collection of (test) particles 
or a single particle with an ensemble of initial conditions.    
  
Now, how does the one-body density $f$ evolve, 
e.g. from one time-like hypersurface to another? We recall two ingredients of 
Liouville's theorem, which will provide the basic tool to answer this question \cite{Goldstein}: 
{\bf i)} Consider a phase-space volume element which is defined by `tracer' 
particles forming its surface; then, due the uniqueness of the Newtonian motion,   
or its relativistic generalization, the number of particles 
inside is constant (in the absence of scattering interactions). 
{\bf ii)} The size of the volume element, being associated with one of Poincar\'e's 
integral invariants (under contact transformations), is also a constant of motion.   
Then, \underline{Liouville's theorem} follows: {\it Iff there are only conservative forces, then 
the phase space density is a constant of motion.}
  
Beginning with this theorem, we derive the evolution equation for $f$ in terms 
of the relativistic proper time $\tau$ as follows: 
\begin{eqnarray}\label{eq1} 
0&=&\frac{\mbox{d}}{\mbox{d}\tau}f(x,p) 
\;\equiv\;\frac{\mbox{d}}{\mbox{d}\tau}\sum_i
\delta^4\left [x_i(\tau )-x\right ]
\delta^4\left [p_i(\tau )-p\right ]
\;\;, \\ [1ex] 
&=&\sum_i\left \{\frac{\mbox{d}x_i^\mu}{\mbox{d}\tau}
\frac{\partial}{\partial x^\mu}
+\frac{\mbox{d}p_i^\mu}{\mbox{d}\tau}
\frac{\partial}{\partial p^\mu}\right \}
\delta^4\left (x_i(\tau )-x\right )
\delta^4\left (p_i(\tau )-p\right )
\nonumber \\ [1ex] \label{eq3} 
&=&\left \{\frac{1}{m}p^\mu 
\frac{\partial}{\partial x^\mu}
+{\cal F}^\mu (x)
\frac{\partial}{\partial p^\mu}\right \}f(x,p) 
\;\;, \end{eqnarray}
where we used the constancy of the density which is represented in 
terms of the particle (index $i$) trajectories $\{ x_i(\tau ),p_i(\tau )\}$ in the 
first line, carried out the differentiation in the second, and employed the 
definition of the four-velocity and the equation of motion in the last, respectively; 
here ${\cal F}^\mu (x)$ denotes the external or selfconsistent internal four-force(s).  
This is the \underline{relativistic Vlasov equation}.

More generally, allowing for scattering of particles into and out of phase space volume 
elements, i.e. $(\mbox{d}/\mbox{d}\tau )f(x,p)\neq 0$, the Vlasov equation is replaced 
by the \underline{Boltzmann equation}: 
\begin{equation}\label{Boltzmann}
\left \{\frac{1}{m}p\cdot\partial_x
+{\cal F}(x)\cdot\partial_p\right \}f(x,p)={\cal C}[f](x,p)  
\;\;, \end{equation}  
where ${\cal C}[f]$ denotes the collision term. It can be derived from 
an analysis of the equations including the two-body densities and generally turns 
out to be a nonlinear functional of the one-body densities.   

A popular set of assumptions for this derivation   
is the following: {\bf i)} Only binary or two-body collisions contribute (in a sufficiently 
dilute system); {\bf ii)} Boltzmann's ``Stosszahlansatz'' according to which the number of 
collisions at $x$ is proportional to $f(x,p)f(x,p')$; {\bf iii)} the distribution function 
varies slowly on the scale of the mean free path, $\lambda_{mfp}|\nabla\log f(x,p)|\ll 1$. 

We remark that while the Vlasov equation is appropriate for systems 
with conservative forces only and, thus, describes nondissipative phenomena, 
the Boltzmann equation incorporates dissipative scattering processes, which 
lead to entropy production.   
 
In order to illustrate this, we introduce the simple 
\underline{relaxation time collision term}, which could be derived rigorously as an approximation of a two-body 
scattering term but can also be seen as a phenomenological ansatz 
taking the dissipation into account: 
\begin{equation}\label{Crelax}
{\cal C}[f]\equiv -\frac{1}{\tau_r}(f-f_0) 
\;\;, \end{equation} 
where $\tau_r$ denotes the relaxation time parameter and $f_0$ the equilibrium 
one-body density, towards which the system will relax.  
  
A particular equilibrium solution of the Boltzmann equation with the above collision 
term can be obtained from the \underline{J\"uttner distribution}, 
\begin{equation}\label{Juettner} 
f_0(x,p)\equiv\exp [-\beta (U\cdot p+\mu )]
\;\;, \end{equation} 
where the local parameters $\{\beta\equiv 1/T,U^\mu,\mu\}(x)$ denote the 
inverse temperature, flow four-velocity, and chemical potential, respectively. 
Instead of the exponential `Boltzmann factor' one may also use a Fermi-Dirac 
or Bose-Einstein distribution, depending on the nature of the considered particles. 
  
In the example of a plasma of charged particles, including a static homogeneous neutralizing 
background, the Lorentz force is ${\cal F}^\mu (x)\equiv F^{\mu\nu}j_\nu =(e/m)F^{\mu\nu}p_\nu$, 
in terms of the field strength tensor $F^{\mu\nu}$ of external and/or 
selfconsistently generated internal electromagnetic fields and the particle electric current $j^\mu$. 
Inserting this together with the J\"uttner distribution, i.e. $f=f_0$, into the 
Boltzmann Eq.\,(\ref{Boltzmann}) together with Eq.\,(\ref{Crelax}), we obtain the equation: 
\begin{equation}\label{Juettner1} 
\frac{\partial}{\partial x_\mu}[\beta (U\cdot p+\mu )]+e\beta F^{\mu\nu}U_\nu =0 
\;\;, \end{equation} 
which constrains the parameters of J\"uttner distribution. Thus, the simplest solution 
indeed is the \underline{global equilibrium distribution} with a constant temperature, a global 
rest frame, and where the gradient of the chemical potential compensates the electric field 
$F^{\mu 0}$, $-\partial_x^\mu \mu =eF^{\mu 0}$.   

\subsection{Relation to relativistic hydrodynamics}
Turning to the observables to be described by the phase space 
density $f$ introduced in Eq.\,(\ref{dP}), we define the \underline{particle 
(mass, charge) four-current}, 
\begin{equation}\label{N}
N^\mu (x)\equiv\int\mbox{d}^4p\;p^\mu f(x,p)
\;\;, \end{equation} 
and the \underline{energy-momentum tensor}, 
\begin{equation}\label{T}
T^{\mu\nu}(x)\equiv\int\mbox{d}^4p\;p^\mu p^\nu f(x,p)
\;\;, \end{equation} 
as the first and second moment of the distribution function, respectively; it is important to recall 
here that we assume $f$ to implicitly contain the usual on-shell constraint, i.e. 
the factors $\Theta (p^0)\delta (p^2-m^2)$, as we discussed. 
  
These are the main quantities of interest in the hydrodynamic description 
of matter, where one integrates out the momentum space information contained in $f$ \cite{Csernai}.    
   
Indeed, it is straightforward to show that $N^\mu$ and $T^{\mu\nu}$ obey the 
appropriate continuity equations related to \underline{mass (or charge) and four-momentum 
conservation}. To begin with, using the Boltzmann Eq.\,(\ref{Boltzmann}) together with 
Eq.\,(\ref{Crelax}), we obtain: 
\begin{equation}\label{Nconserv}
\partial_\mu N^\mu =\int\mbox{d}^4p\;p\cdot\partial_x f 
=-m\int\mbox{d}^4p\left ({\cal F}\cdot\partial_pf+\frac{f-f_0}{\tau_r}\right )=
-\frac{m}{\tau_r}\int\mbox{d}^4p\;(f-f_0)\equiv -\partial_\mu\delta N^\mu
\;\;, \end{equation}
where we find a dissipative contribution $\delta N^\mu$ on the right-hand side, which vanishes only 
if the ordinary density equals the equilibrium density determined by $f_0$. Similarly, one obtains: 
\begin{equation}\label{Tconserv}
\partial_\mu T^{\mu\nu}=-m\int\mbox{d}^4p\;p^\nu\left ({\cal F}\cdot\partial_pf+\frac{f-f_0}{\tau_r}\right )
=m{\cal F}^\nu (\int\mbox{d}^4p\;f)-\frac{m}{\tau_r}(N^\nu -N_0^\nu )
\;\;, \end{equation}
where the second term on the right-hand side is related to a dissipative contribution 
$-\partial_\mu\delta T^{\mu\nu}$ to the 
energy-momentum tensor, while the first term presents the external or 
selfconsistent force density acting on the system; $N_0^\mu$ is defined like $N^\mu$, however, with $f$ replaced 
by $f_0$. 
   
More generally, based on the Chapman-Enskog method, standard forms of the dissipative terms 
can be constructed incorporating the \underline{transport coefficients} of shear and bulk viscosity, 
heat conductivity, particle production/annihilation, and diffusion \cite{Csernai,ERT01}. 

Finally, we consider the entropy. It is an important quantity not only 
in equilibrium thermodynamics, but can be used, for example, to characterize the bulk 
properties of matter produced in high-energy collisions. In particular,  
it can be related more or less directly to the observed particle multiplicities 
\cite{Bj,QMs,Csernai}. Here we define the \underline{entropy four-current}: 
\begin{equation}\label{S}
S^\mu (x)\equiv -\int\mbox{d}^4p\;p^\mu f(x,p)\ln [f(x,p)/f_0(x,p)]
\;\;. \end{equation} 
Calculating as before, we obtain the entropy production formula: 
\begin{equation}\label{Sconserv} 
\partial_x\cdot S=\frac{m}{\tau_r}\int\mbox{d}^4p\;
(\ln [f/f_0]+1)(f-f_0)
\;\; \end{equation} 
Since $(\ln x +1)(x-1)\geq 0$, for all $x\geq 0$, we recover  
\underline{Boltzmann's ``H-theorem''}: 
\begin{equation}\label{Htheorem}
\partial_x\cdot S\geq 0
\;\;, \end{equation} 
expressing a positive entropy production which vanishes only in equilibrium, when $f=f_0$.  

This completes our overview of classical relativistic transport theory and some 
of its ramifications and we turn to quantum mechanics next.   

\section{Quantum mechanical distribution functions \\ 
- the density matrix and the Wigner function}
In order to motivate the necessity for a density matrix formulation 
in quantum mechanics, we recall Feynman's famous division \cite{Feynman,Kim99}, 
\begin{equation}\label{RestofU}
{\cal U}niverse\;\equiv\;{\cal S}ystem\;+\;{\cal R}est
\;\;, \end{equation} 
which says that doing physics and paying attention to the system requires 
precisely to separate off the rest of the Universe, i.e. the `environment'  
of the system in modern terminology. 
  
Following the presentation in Ref.\,\cite{Feynman}, we introduce generic coordinates 
$x$ and $y$ for the description of ${\cal S}$ and ${\cal R}$, respectively. 
Assuming the existence of corresponding complete sets of states or wave 
functions, $\{\phi_i(x)\}$ and $\{\psi_j(y)\}$, 
the most general state vector and normalized wave function 
of ${\cal U}$ can be expanded, respectively, as: 
\begin{equation}\label{statewf} 
|\Psi\rangle =\sum_{i,j}c_{ij}|\phi_i\rangle |\psi_j\rangle 
\;\;,\;\;\;\Psi (x,y)=\sum_{i,j}c_{ij}\langle y|\phi_i\rangle\langle x|\psi_j\rangle 
\equiv\sum_i c_i(y)\phi_i(x) 
\;\;, \end{equation}
with complex expansion coefficients $c_{ij}$ and functions $c_i(y)$. 
  
Considering an operator $\hat A$ which acts on ${\cal S}$ only, it is  
given in terms of its matrix elements by: 
\begin{equation}\label{A} 
\hat A\equiv\sum_{ii'j}A_{ii'}|\phi_i\rangle |\psi_j\rangle\langle\psi_j|\langle\phi_{i'}|  
\;\;, \end{equation} 
i.e., it acts as a projector on ${\cal R}$. Following the rules, we obtain the 
expectation value of $\hat A$ in the state $|\Psi\rangle$ of Eq.\,(\ref{statewf}): 
\begin{equation}\label{Aexp} 
\langle\hat A\rangle\equiv\langle\Psi |\hat A|\Psi\rangle 
=\sum_{i'ij'}A_{i'i}C_{ij'}C^\ast_{i'j'}\equiv\sum_{i'i}A_{i'i}\rho_{ii'}=\mbox{Tr}\;\hat A\hat\rho 
=\mbox{Tr}\;\hat\rho\hat A
\;\;. \end{equation} 
It follows from its definition here that the density operator $\hat\rho$, with 
the density matrix elements identified as $\rho_{ii'}\equiv\langle\phi_i|\hat\rho |\phi_{i'}\rangle$, 
is a hermitean operator. 
  
Therefore, we may introduce a complete orthonormal basis $\{ |i\rangle\}$ diagonalizing $\hat\rho$: 
$\hat\rho =\sum_iw_i|i\rangle\langle i|\;$,  
with real eigenvalues $w_i$. Furthermore, choosing $\hat A=1_{\cal S}\otimes1_{\cal R}$, i.e. the 
identity operator, one finds with the help of Eq.\,(\ref{Aexp}) the sum rule:  
$1=\langle\hat A\rangle =\mbox{Tr}\;\hat\rho =\sum_iw_i\;$. Finally, choosing instead 
$\hat A=|i'\rangle\langle i'|\otimes1_{\cal R}$, a similar calculation yields: 
$w_{i'}=\langle\hat A\rangle =\langle\Psi |\hat A|\Psi\rangle =\sum_j|\langle\psi_j |\langle i'|\Psi\rangle |^2
\geq 0\;$. 
  
Abstracting from the present example, it is postulated that {\it any quantum mechanical system} is to 
be described by a \underline{hermitean density operator}, 
\begin{equation}\label{rho}
\hat\rho =\sum_iw_i|i\rangle\langle i|
\;\;, \end{equation} 
where $\{ |i\rangle\}$ forms a complete orthonormal set, the real expansion coefficients 
are non-negative, $w_i\geq 0$, and fullfill a normalization condition, $\sum_iw_i=1$. -- We 
remark that the density matrix was first introduced by von\,Neumann in 1932 in his by now famous 
book \cite{JvN}. --  
Observables in particular and expectation values of operators in general are 
always to be calculated by: 
\begin{equation}\label{Aexp1}
\langle\hat A\rangle =\mbox{Tr}\;\hat\rho\hat A 
=\sum_iw_i\langle i|\hat A|i\rangle
\;\;. \end{equation} 
Consequently, the coefficients $w_i$ are interpreted as describing the probability to find 
the system in the state $|i\rangle$. 
  
As a matter of nomenclature, one distinguishes \underline{pure and mixed states} of a system, 
the former being defined by a density operator which is a projector, 
$\hat\rho =\hat\rho^2\Longleftrightarrow (w_{i^\ast}=1,\;\mbox{all other}\; w_i=0)$, and the latter 
comprising all other cases. -- We recall the case of a system in contact with a heat 
bath, which leads to the mixed state density operator of \underline{thermal equilibrium}: 
\begin{equation}\label{thermalrho} 
\hat\rho (\beta )=Z^{-1}(\beta )\sum_n e^{-\beta E_n}|E_n\rangle\langle E_n|
\;\;,\;\;\; Z(\beta )\equiv\sum_ne^{-\beta E_n}
\;\;, \end{equation} 
where $\{ |E_n\rangle\}$ denotes the complete set of normalized energy eigenstates of the system and $Z$ is 
its partition function necessary to normalize the exponential `Boltzmann factor'. All 
canonical thermodynamical relations can be easily derived from this density operator, 
see, for example, \cite{Feynman}.  
  
Finally, we may address also in the present context the question 
how the system, i.e. its density operator, evolves in time. 
This is easily answered by expanding the eigenstates $\{ |i\rangle\}$ of $\hat\rho$ in terms of 
the energy eigenstates, 
\begin{equation}\label{iexpans}
|i(0)\rangle\equiv |i\rangle =\sum_n|E_n\rangle\langle E_n|i(0)\rangle
\;\;, \end{equation}  
which implies according to the Schr\"odinger equation: $|i(t)\rangle =\exp (-i\hat Ht)|i(0)\rangle$, given 
the Hamiltonian $\hat H$. Then, the \underline{evolution of $\hat\rho$} follows: 
\begin{equation}\label{rhoevolv} 
\hat\rho (t)=e^{-i\hat Ht}\hat\rho (0)e^{+i\hat Ht}
\;\;. \end{equation} 
Equivalently, we obtain: 
\begin{equation}\label{rhoequ} 
\frac{\mbox{d}}{\mbox{d}t}\hat\rho (t)=i[\hat\rho (t),\hat H]
\;\;, \end{equation} 
which differs from the usual operator evolution equation in quantum mechanics 
by an additional minus sign on the right-hand side. Both signs, of course, are consistent, 
since we have: 
\begin{equation}\label{expectevolv} 
\langle\hat A\rangle_{(t)}\equiv\mbox{Tr}\;\hat\rho (t)\hat A
=\mbox{Tr}\;\hat\rho (0)e^{+i\hat Ht}\hat Ae^{-i\hat Ht}\equiv\langle\hat A(t)\rangle 
\;\;, \end{equation} 
using the cyclicity of the trace. 

The equations (\ref{rhoequ}) and (\ref{expectevolv}) present the problem of quantum 
transport theory in its most compact form. In particular, Sections 4--6 are devoted 
to explorations of various more detailed forms of these abstract results and their 
applications. 
  
In order to connect the density matrix formalism with the classical transport 
theory, we turn to the density matrix in coordinate and momentum representation.  

It helps to visualize the following by imagining a single particle to be 
described quantum mechanically. Then, in the \underline{coordinate representation}, 
using corresponding single-particle wave functions, we 
have the density matrix elements:
\begin{equation}\label{rhoxx}
\rho (x',x)=\sum_iw_i\langle x'|i\rangle\langle i|x\rangle 
=\sum_iw_i\phi_i(x')\phi_i^\ast (x) 
\;\;, \end{equation} 
see Eq.\,(\ref{rho}). Then, we calculate immediately the 
probability to find the particle at $x$: 
\begin{equation}\label{Px}
P(x)\equiv\rho (x,x)=\sum_iw_i|\phi_i(x)|^2 
\;\;, \end{equation} 
in agreement with the standard rules of quantum mechanics. Similarly, 
in \underline{momentum representation},
\begin{equation}\label{rhopp}
\rho (p',p)=\sum_iw_i\langle p'|i\rangle\langle i|p\rangle 
=\sum_iw_i\phi_i(p')\phi_i^\ast (p) 
\;\;, \end{equation} 
which yields: 
\begin{equation}\label{Pp}
P(p)\equiv\rho (p,p)=\sum_iw_i|\phi_i(p)|^2 
\;\;, \end{equation} 
i.e. the probability to find the particle with momentum $p$. 
  
Now, considering a function $O$ -- representing some observable, for example -- which is defined over phase space, 
we obtain its average value by calculating: 
\begin{equation}\label{classav}
\overline{O}\equiv\int\mbox{d}x\mbox{d}p\;O(x,p)f(x,p) 
\;\;, \end{equation} 
involving the one-body probability density function $f$. The question then arises, whether 
there exists a corresponding quantum mechanical density function which yields 
the expectation value of operators in the form of phase space integrals,  
generalizing Eq.\,(\ref{classav}). One possible answer is provided by the 
\underline{Wigner function} \cite{Feynman,Kim99,Carruthers}: 
\begin{equation}\label{Wignerf}
W(x,p)\equiv\int\mbox{d}y\;e^{ipy/\hbar}\rho (x+y/2,x-y/2)
\;\;, \end{equation} 
i.e. a particular Fourier transform of the density matrix in coordinate space. 
We remark that the momentum appearing as an argument of $W$ is the conjugate 
variable to the relative coordinate separating the pair of wave functions which enter $\rho$, 
cf. Eq.\,(\ref{rhoxx}); we also indicate here explicitly the $\hbar$-dependence 
arising in the phase factor. 
  
Furthermore, even though the Wigner function presents a real distribution, 
due to $W(x,p)=W^\ast (x,p)$, it may oscillate and indeed does so in most cases. 
Nevertheless, we easily obtain the following results which match their 
classical counterparts: 
\begin{eqnarray}\label{Pxint}
\int\frac{\mbox{d}p}{2\pi\hbar}\;W(x,p)&=&\rho (x,x)=P(x)
\;\;, \\ [1ex]
\label{Ppint} 
\int\mbox{d}x\;W(x,p)&=&\rho (p,p)=P(p)
\;\;. \end{eqnarray}  
This implies for the \underline{expectation values} of functions of 
operators: 
\begin{eqnarray}\label{Oxexp}
\langle O(\hat x)\rangle &=&\mbox{Tr}\;\hat\rho O(\hat x)
=\int\frac{\mbox{d}x\mbox{d}p}{2\pi\hbar}\;O(x)W(x,p)
\;\;, \\ [1ex]
\label{Opexp}
\langle O(\hat p)\rangle &=&\mbox{Tr}\;\hat\rho O(\hat p)
=\int\frac{\mbox{d}x\mbox{d}p}{2\pi\hbar}\;O(p)W(x,p)
\;\;, \end{eqnarray} 
which should be compared to Eq.\,(\ref{classav}).   
  
However, besides the fact that the Wigner function generally is 
not positive definite, the expected limitation of the classical/quantum 
correspondence of various (probability) density functions also shows up, when one 
considers operators of the form $O(\hat x,\hat p)$. In this case, 
operator ordering becomes an issue and the above formulae have to 
be generalized with care. On the other hand, as we shall see shortly in  
Section 4, the appropriate semiclassical expansion of the Wigner function  
transport equation does yield the transport equation for 
the classical probability density function, 
which we discussed in Section 2.  

\section{Transport theory for quantum fields}
As an example of a field theory we choose the O(4) {\it linear $\sigma$-model}, 
which has a long history of applications in various phenomenological contexts. 
Recently it has been argued by Wilczek that it represents the QCD chiral 
order parameter for $n_f=2$ massless quark flavors \cite{Wilczek92}. Furthermore, 
it has been demonstrated by nonperturbative calculation that this model possesses a 
second order finite temperature phase transition between the spontaneously broken and 
symmetry restored phases \cite{Wetterich93}. Most interestingly, the effective mass 
and coupling go to zero at the phase transition with well-determined critical exponents. 
This may influence the hydrodynamic behavior of matter described by this model in  
interesting ways, which we presently study. 

Here, our aim is to derive the quantum transport equations in 
the Hartree approximation. In particular, we will illustrate how an appropriate 
Wigner function can be introduced and how field theory aspects are related to 
our earlier considerations of classical transport theory. 

To begin with, the O(4)-invariant 
$\sigma$-model action is defined by:
\begin{equation}\label{sigaction}  
S[\vec\phi ]\equiv\int\mbox{d}^4x\left (\frac{1}{2}(\partial\vec\phi )^2
-\frac{1}{2}\mu^2\vec\phi^2-\frac{1}{4!}\lambda (\vec\phi^2)^2\right )
\;\;, \end{equation} 
with $\vec\phi\equiv (\phi_1,\phi_2,\phi_3,\phi_4)$ and 
where the mass parameter is chosen with the `wrong' sign, i.e. $\mu^2<0$. The 
corresponding potential, 
$V(\vec\phi )\equiv\frac{1}{2}\mu^2\vec\phi^2+\frac{1}{4!}\lambda (\vec\phi^2)^2$, 
is of the so-called `Mexican hat' form which leads to 
spontaneous symmetry breaking at the tree level.

We parametrize the four-component vector $\vec\phi$ in terms of 
three-component `pion' and one-component `sigma' fields, $\vec\phi\equiv (\vec\pi ,\sigma')$, 
at each space-time point. Shifting the $\sigma'$-field by the vacuum expectation value 
$\sigma_0$ of $\vec\phi$, which is determined by the minimum of the potential, 
\begin{equation}\label{minV} 
\frac{\mbox{d}V(\vec\phi )}{\mbox{d}\vec\phi}=(\mu^2+\frac{1}{3!}\lambda\vec\phi^2)
\vec\phi =0\;\;\;\Longrightarrow\;\;\;\; |\sigma_0|=\sqrt{-6\mu^2/\lambda}
\;\;, \end{equation} 
we define the fluctuation field $\sigma\equiv\sigma'-\sigma_0$. 

The effect of this parametrization combined with the shift of one component by the 
vacuum expectation value is easily seen in the corresponding 
\underline{Heisenberg operator} equations of motion, 
\begin{eqnarray}\label{PiOp}
&\;&\partial^2\vec\pi +m_\pi^2\vec\pi+\frac{\lambda}{3!}\vec\pi^2\vec\pi =0
\;\;, \\ [1ex] \label{SigOp} 
&\;&\partial^2\sigma +m_\sigma^2\sigma +\frac{\lambda}{2}\sigma_0\sigma^2+\frac{\lambda}{3!}\sigma^3 =0
\;\;, \end{eqnarray} 
which are obtained by varying the action $S[\vec\phi ]$ with 
respect to $\vec\phi$, introducing parametrization and shift, and considering the 
fields as quantum field operators. Here we introduced the effective masses: 
\begin{equation}\label{effmass}
m_\pi^2\equiv\frac{\lambda}{3!}(\sigma^2+2\sigma_0\sigma )\;\;,\;\;\;
m_\sigma^2\equiv\frac{\lambda}{3!}(\vec\pi^2+2\sigma_0^{\;2})
\;\;. \end{equation}. 
We observe that for $\sigma\rightarrow 0$ we have $m_\pi^2\rightarrow 0$, while 
$m_\sigma^2\rightarrow\lambda\sigma_0^{\;2}/3$, yielding     
one massive `radial' mode together with three massless `Goldstone modes',  
in accordance with the Goldstone theorem.  

Taking the expectation value Eqs.\,(\ref{PiOp}) and (\ref{SigOp}), 
we observe that one-point functions, such as the \underline{mean field} $\bar\sigma (x)\equiv\langle\sigma (x)\rangle$, 
generally are coupled to two-point functions, such as  $\lim_{x'\rightarrow x}\langle\sigma(x)\sigma(x')\rangle$, 
and higher $n$-point functions. The coincidence limit of these \underline{Wightman functions} 
produces divergences necessitating a renormalization procedure, 
which we will discuss at the end of this section. 

In order to solve the resulting equations, 
one has to specify the density operator which enters the expectation values, e.g. $\langle\sigma\rangle\equiv\mbox{Tr}\;\rho\sigma$, omitting the operator signs used previously, cf. Section 3. 
This can be done, for example, on a fixed time-like hypersurface, which will be 
demonstrated for fermions in Section 6. Furthermore, 
we need to derive equations of motion for the higher $n$-point functions, since taking expectation values of nonlinear operator equations automatically generates an infinite hierarchy of equations, similarly as the Schwinger-Dyson equations for propagators or the BBGKY hierarchy in classical transport theory.     
  
The simplest nonperturbative truncation of the hierarchy of Wightman function equations 
is produced by the \underline{Hartree approximation}. It consists in factorizing
the $n$-point functions into products of one- or two-point functions, properly taking into 
account all possible factorizations. For example, we obtain:
\begin{equation}\label{fact} 
\langle\sigma_1\sigma_2\sigma_3\rangle =
\bar\sigma_1\langle\underline\sigma_2\underline\sigma_3\rangle 
+\bar\sigma_2\langle\underline\sigma_1\underline\sigma_3\rangle
+\bar\sigma_3\langle\underline\sigma_1\underline\sigma_2\rangle
+\bar\sigma_1\bar\sigma_2\bar\sigma_3
\;\;, \end{equation}
where the subscripts refer to different space-time points. 
Note that the expectation value of any odd power of the proper 
quantum field vanishes, since we define $\underline\sigma\equiv\sigma -\bar\sigma$. -- 
This approximation is known to be equivalent to summing all iterated bubbles (`superdaisies') in 
the Feynman diagram calculation of the vacuum effective action in a $\phi^4$-model \cite{CJT}. 

Here we will furthermore assume that cross terms vanish, e.g.  
$\langle\vec\pi\sigma\rangle =0$. It turns out to be consistent with this assumption to set  
$\langle\vec\pi\rangle =0$, since the classical pion field obtains no source term. In distinction, 
the Eq.\,(\ref{SigOp}) yields the Klein-Gordon type 
\underline{mean field equation}: 
\begin{equation}\label{SigMF}
\left (\partial^2+m_\sigma^2 \right .
+\frac{\lambda}{2}(\frac{1}{3}\bar\sigma^2 +\sigma_0\bar\sigma + \left. \langle \underline{\sigma}^2\rangle \right )
\bar\sigma
+\frac{\lambda}{2}\sigma_0\langle\underline{\sigma}^2\rangle 
\equiv\left(\partial^2+m_\sigma^2+\delta m^2\right )\bar\sigma +J=0
\;\;, \end{equation}
where the two-point function adds to the nonlinear classical force term shifting the effective 
$\sigma$-mass, which contains $\langle\vec\pi^2\rangle$, 
and to a source term, $\delta m^2$ and  
$J$, respectively. For a homogeneous system, for $\delta m^2\gg m_\sigma^2$, and for sufficiently large $\langle\underline\sigma^2\rangle$, the 
mean field is $\langle\sigma\rangle =-\sigma_0$, indicating  
symmetry restoration with $\langle\sigma'\rangle =0$. 

Multiplying Eq.\,(\ref{SigOp}) 
by one more power of the field operator from the left, before applying the Hartree approximation as before, 
we obtain the \underline{two-point function equation}: 
\begin{equation}\label{Sig2P}
\left (\partial_2^{\;2}+m_{\sigma ,2}^2+\lambda\sigma_0\bar\sigma_2+
\frac{\lambda}{2}(\bar\sigma_2^{\;2}+\langle\underline\sigma_2^{\;2}\rangle) 
\right )\langle\underline\sigma_1\underline\sigma_2\rangle 
 =0
\;\;, \end{equation}
where subscripts ``1,2'' refer to two different space-time points. 
Here we also used Eq.\,(\ref{SigMF}) at point ``2'', multiplied by $\bar\sigma_1$, in order to 
simplify Eq.\,(\ref{Sig2P}) considerably. A similar equation 
follows from Eq.\,(\ref{PiOp}) for the `pion' modes. 
  
The next step consists in defining a suitable \underline{Wigner operator}, 
\begin{equation}\label{WigOp}
W_{ab}(x,p)\equiv\int\frac{\mbox{d}^4y}{(2\pi )^4}e^{-ip\cdot y}
\underline\Phi_a(x+\frac{1}{2}y)\underline\Phi_b(x-\frac{1}{2}y)
\;\;, \end{equation} 
where $\vec\Phi\equiv (\vec\pi ,\sigma)$ and $\underline\Phi\equiv\Phi -\bar\Phi$. 
This should be compared to the quantum mechanical Wigner function 
introduced previously, Eq.\,(\ref{Wignerf}). We remark that it might be useful for 
some applications not involving vacuum properties to normal-order the field operators 
in the definition of $W$, see, for example, Section 6.

Illustrating the usefulness of $W$, we write down the energy-momentum tensor for the O(4)-model: 
\begin{eqnarray}\label{TmunuPhi} 
T_{\mu\nu}&=&\left\langle (\partial_\mu\vec\Phi )\cdot (\partial_\nu\vec\Phi )-g_{\mu\nu}
[\frac{1}{2}(\partial\vec\Phi )^2-\frac{1}{2}\mu^2\vec\Phi^2-\frac{1}{4!}\lambda (\vec\Phi^2)^2]\right\rangle 
\\ [2ex] 
&=&\int\mbox{d}^4p\;\left (p_\mu p_\nu+\frac{1}{4}\partial_{x^\mu}\partial_{x^\nu}
-\frac{1}{2}g_{\mu\nu}(p^2+\frac{1}{4}\partial_x^{\;2})\right )\langle W_{aa}(x,p)\rangle 
\nonumber \\ [1ex] \label{TmunuW}
&\;&+g_{\mu\nu}\int\mbox{d}^4p\;\left (\frac{1}{2}\mu^2\langle W_{aa}(x,p)\rangle 
+\frac{1}{4!}\lambda\int\mbox{d}^4p'\;\langle W_{aa}(x,p)W_{bb}(x,p')\rangle\right )
\;\;, \end{eqnarray}
to which must be added the purely classical terms plus mean field dependent interaction terms ($\propto\lambda$) 
involving $W$, 
which can all be further evaluated in Hartree approximation. 
   
Heading for the transport theory, we will express Eq.\,(\ref{Sig2P}) in terms of the Wigner operator.  
We introduce the abbreviation: 
\begin{equation}\label{M}
{\cal M}^2(x)\equiv m_\sigma^2(x)+\lambda\sigma_0\bar\sigma (x)+
\frac{\lambda}{2}\left (\bar\sigma^{\;2}(x)+\langle\underline\sigma^{\;2}(x)\rangle\right )
\;\;. \end{equation}  
The two-point functions contained here, $\langle\underline{\vec\pi}^2\rangle$ in $m_\sigma^2$ and 
$\langle\underline\sigma^2\rangle$, respectively, can be rewritten using the Wigner operator, e.g.:  
$\langle\underline{\vec\sigma}^2(x)\rangle =\int\mbox{d}^4p\langle W_{\sigma\sigma}(x,p)\rangle$. 
Then we obtain instead of Eq.\,(\ref{Sig2P}):  
\begin{equation}\label{Weq}
\left (\frac{1}{4}\partial_x^{\;2}-p^2+ip\cdot\partial_x
+\exp (-\frac{i}{2}\partial_x\cdot\partial_p){\cal M}^2(x)\right )\langle W_{\sigma\sigma}(x,p)\rangle 
=0
\;\;, \end{equation} 
where the $x$-derivative in the exponential acts only on ${\cal M}^2$. In the derivation of this result, 
and similarly for the `pions',  
one frequently makes use of suitable partial integrations under the momentum integral from $W$, as well 
as expressing the shifted argument of ${\cal M}^2$ by a translation operator giving rise to the 
exponential \cite{EH,VGE87}.  

The final step consists in adding to/subtracting from the complex Eq.\,(\ref{Weq}) its adjoint. 
This yields the \underline{transport equation}: 
\begin{equation}\label{Wtrans} 
\left (p\cdot\partial_x
-\frac{1}{\hbar}\sin (\frac{\hbar}{2}\partial_x\cdot\partial_p){\cal M}^2(x)\right )
\langle W_{\sigma\sigma}(x,p)\rangle 
=0
\;\;, \end{equation} 
together with a generalized \underline{mass-shell constraint}: 
\begin{equation}\label{Wmass} 
\left (p^2-\frac{\hbar^2}{4}\partial_x^{\;2}
-\cos (\frac{\hbar}{2}\partial_x\cdot\partial_p){\cal M}^2(x)\right )\langle W_{\sigma\sigma}(x,p)\rangle 
=0
\;\;, \end{equation} 
where the appropriate powers of $\hbar$ are reinserted. We stress that the  
covariant transport equation alone does not suffice to determine the dynamics.     
   
A systematic semiclassical expansion of Eqs.\,(\ref{Wtrans})-(\ref{Wmass}) in powers of $\hbar$ 
becomes feasible now. To leading order we obtain: 
\begin{eqnarray}\label{WtransL} 
\left (p\cdot\partial_x
-\frac{1}{2}\partial_x\cdot\partial_p{\cal M}^2(x)\right )
\langle W_{\sigma\sigma}(x,p)\rangle 
&=&0
\;\;, \\ [1ex] 
\label{WmassL} 
\left (p^2
-{\cal M}^2(x)\right )\langle W_{\sigma\sigma}(x,p)\rangle 
&=&0
\;\;, \end{eqnarray} 
which are indeed of the form of a classical Vlasov and mass-shell equation, respectively, 
discussed in Section 2.1. Note, however, that the effective mass or \underline{effective potential}, 
appearing here contains contributions from the mean field, which have to be determined 
selfconsistently from Eq.\,(\ref{SigMF}), and `selfenergy' terms, which 
we will discuss shortly.  
  
We remark that effects of the higher-order \underline{$\hbar$-corrections} to Eq.\,(\ref{WtransL}) and 
(\ref{WmassL}) are largely unexplored, since numerically the corresponding higher derivatives 
lead to strong instabilities. This provides part of the motivation to develop an analytical  
approach solving the exact quantum transport equations, the first step of which is described 
in Section 6. 
  
Furthermore, we observe that the equations derived here do not yield any collision terms. 
This was to be expected, since the \underline{Hartree approximation}, which is essentially equivalent 
to the Gaussian approximation in the Schr\"odinger functional approach, leads to an effective 
`quantum' Hamiltonian evolution which is \underline{nondissipative} \cite{I95}. Only an improved 
treatment of correlation terms, such as the last term in Eq.\,(\ref{TmunuW}), will go beyond 
the collisionless Vlasov dynamics obtained here, see, for example, Refs.\,\cite{Cooper98,Blaizot99}.   

Finally, we turn to the `selfenergy' terms mentioned above and to the divergences caused by them, 
in particular. These terms involve integrals of the kind $\int\mbox{d}^4p\langle W_{\sigma\sigma}(x,p)\rangle$.  
Let us consider the simplest case of a density operator projecting on the vacuum state, 
$\rho_{vac} =|0\rangle\langle 0|$. Then, for a generic free scalar field, the \underline{vacuum Wigner function} is:
\begin{equation}\label{Wvac} 
W_{vac}(p)\equiv\mbox{Tr}\;\rho_{vac}W(x,p)=\langle 0|W(x,p)|o\rangle =(2\pi )^{-3}\delta (p^2-m^2) 
\;\;, \end{equation} 
where $m$ denotes its {\it physical} mass, and the calculation proceeds, for example, by expanding the field operators 
appearing in the definition of the Wigner operator in terms of creation and annihilation operators. 
In this case, we obtain: 
\begin{equation}\label{selfenergy} 
\int\mbox{d}^4p\;W_{vac}(p)=\int\frac{\mbox{d}^3p}{(2\pi )^3\omega_p}\equiv I_m 
\;\;, \end{equation} 
with $\omega_p\equiv (\vec p^2+m^2)^{1/2}$. This `selfenergy' integral is \underline{quadratically divergent}.  
 
Since divergences arise from field operators at coinciding points and, as a first approximation, independently  
of which point it is, we rewrite Eq.\,(\ref{Wmass}) accordingly, using Eqs.\,(\ref{effmass}) and (\ref{M}): 
\begin{eqnarray}\label{WmassR1} 
0&=&(-p^2+m_{bare}^2
+\frac{\lambda}{2}I_m
+\frac{\lambda}{2}\int\mbox{d}^4k\left\{ W(x,k)-W_{vac}(k)\right\})
(W(x,p)-W_{vac}(p)+W_{vac}(p))
\\ [1ex] \label{WmassR2}
&\equiv&\left (-p^2+m_r^{\;2}+\frac{\lambda}{2}\int\mbox{d}^4k\;W_r(x,k)\right )(W_r(x,p)+W_{vac}(p))
\;\;, \end{eqnarray} 
with $W(x,p)\equiv\langle W_{\sigma\sigma}(x,p)\rangle$ and  
the Lagrangian `bare' mass $m_{bare}^2\equiv\frac{\lambda}{3}\sigma_0^{\;2}$, 
cf. Eq.\,(\ref{effmass}). Here we left out all `pion' and mean field contributions to 
keep things transparent. These can be added following  
the same strategy, while the gradient corrections would force us to keep $W(x,p)$ unrenormalized 
at first sight. 

Note that we added and subtracted 
suitably the divergent term in 
Eq.\,(\ref{WmassR1}), thus introducing in Eq.\,(\ref{WmassR2}) 
the renormalized mass and Wigner function, $m_r$ and $W_r$, respectively. 
We complete the \underline{renormalization} by identifying the physical mass: 
\begin{equation}\label{mphys}
m^2\equiv m_r^{\;2}+\frac{\lambda}{2}\int\mbox{d}^4k\;W_r(x,k) 
\;\;. \end{equation} 
With this, the Eq.\,(\ref{WmassR2}) 
assumes a simple form,  
$(p^2-m^2)W_r(x,p)=0$, indicating that $W_r$ is (in our first approximation) 
an on-shell distribution: 
\begin{equation}\label{WRos}
W_r(x,p)=\delta (p^2-m^2)f(x,p)
\;\;, \end{equation} 
such as a Bose-Einstein distribution with (weakly) $x$-dependent parameters, cf. Section 2.1.
Neglecting this dependence altogether and assuming a simple thermal distribution turns 
Eq.\,(\ref{mphys}) into a mass-gap equation with $m_r^{\;2}$ as input parameter and 
a resulting temperature dependent mass $m^2(T)$.   

A mass renormalization as performed here is sufficient to render the mean field Eq.\,(\ref{SigMF}) and 
the operator of two-point function Eq.\,(\ref{Sig2P}) finite. The gradient corrections in the transport 
and generalized mass-shell constraint, however, indicate that the renormalized Wigner (or equivalently the 
two-point) function has to be be renormalized separately, subtracting as above, and thus closing the set of equations.   
Generally, the situation gets more complicated, when the $x$-dependence of 
$W_r$ has to be taken serious and, thus, the physical mass becomes space-time dependent, e.g. in the 
case of a system which evolves from strongly inhomogeneous initial conditions.  
We will not address these issues here, but turn to a related physical effect in the next section.
        
\section{Particle production by classical fields}
The color string, rope, or flux tube models are widely used in phenomenological 
descriptions of particle production, in high-energy nuclear collisions in particular \cite{QMs}. 
The basis here is Schwinger's nonperturbative calculation of vacuum decay due 
to charged particle production in {\it constant} and {\it homogeneous} external electric fields 
\cite{Schwinger,IZ}. Considering the inhomogeneous and rapid evolution of the system 
during a heavy-ion collison, the basic assumptions of this picture are questionable. 

Since an appropriate generalization of Schwinger's result is still not available, 
one may as well consider the perturbative evaluation of particle production, 
however, for arbitrarily varying fields. This was originally discussed in Ref.\,\cite{I88} 
for QCD in abelian dominance approximation. There, also the ensuing modification of 
semiclassical transport equations, cf. Section 4, and the related vacuum polarization 
current were obtained. Here we briefly recall part of this calculation which is based 
on an $O(g^2)$ solution of the quark transport equation in external color fields. 

Simplifying the notation, we consider electrically charged Dirac fermions (mass $m$) in arbitrary 
electromagnetic fields. We define the Fourier transformed \underline{vacuum Wigner function}:  
\begin{eqnarray}\label{WFvac}
f_{vac}(q,p)&\equiv&\int\mbox{d}^4x\mbox{d}^4y\;e^{iq\cdot x}e^{-ip\cdot y}
\langle 0|\bar\psi (x+\frac{1}{2}y)\psi (x-\frac{1}{2}y)|0\rangle
\\ [1ex] \label{WFvac1} 
&=&-(2\pi )^5\delta^4(q)\delta (p^2-m^2)\theta (-p^0)(\gamma\cdot p+m) 
\;\;, \end{eqnarray}
which is a 4x4 spinor matrix. The final form here follows similarly 
as in the scalar case, Eq.\,(\ref{Wvac}), using a standard expansion of the field operators \cite{IZ}. 
In distinction to the fermion Wigner function introduced in the following Section 6, Eq.\,(\ref{Wigner}), 
where the notation is explained in more detail, we presently do not normal-order the field operators, since we are 
interested to study the response of the vacuum, i.e. the modification of $f_{vac}$, due to external fields.       

Our task is to somehow solve the \underline{full quantum transport equation} which determines the 
fermionic Wigner function. For the present study the transport equation based on the 
linear form of the Dirac equation is most useful \cite{EH,VGE87}. -- The quadratic form yields 
transport equations which are closer to the usual form, such as in the scalar field case studied in Section 4. --  
Presently, following Fourier transformation, we have: 
\begin{equation}\label{ftrans}
\left (\gamma\cdot (p+\frac{1}{2}q)-m\right )f(q,p)=g\int\frac{\mbox{d}^4q'}{(2\pi )^4}\;\gamma\cdot A(q')
f(q-q',p-\frac{1}{2}q')
\;\;, \end{equation} 
where $A^\mu$ denotes the vector potential of the external field. This equation has to be solved 
together with the \underline{constraint}: 
\begin{equation}\label{fconstr} 
f^\dagger(q,p)=\gamma^0f(-q,p)\gamma^0
\;\;, \end{equation} 
which follows from the definition of the fermion Wigner function; equivalently, one may solve 
Eq.\,(\ref{ftrans}) simultaneously with its adjoint.  
  
The perturbative solution in powers of the coupling constant $g$ of Eq.\,(\ref{ftrans}) can 
be found interatively, starting with the zeroth order solution $f_{vac}$ as input on the 
right-hand side. 

Thus, at first order one finds the \underline{induced vacuum current}, 
$J_{(1)}^\mu (q)=\int\mbox{d}^4p\mbox{Tr}\gamma^\mu f_{(1)}^\mu (q,p)$, which 
can be written explicitly involving the first order (one-loop) QED vacuum 
polarization tensor and reproduces the known result \cite{IZ,I88}. 
Furthermore, since there is a gauge invariant version of Eq.\,(\ref{ftrans}) based on 
a modified definition of the Wigner function \cite{EH,VGE87,I88}, this type of 
calculation might allow to calculate the polarization tensor in a manifestly 
gauge invariant way.    

For our present purposes, the second order solution of the  
equations is most interesting. The main algebraic complication arises from Eq.\,(\ref{fconstr}), which 
has to be satisfied order by order. We do not give the lengthy expressions here, but turn to the 
calculation of the \underline{spectrum} of the produced particles. 
We have to use reduction formulae, in order to relate the `in' field operators defining the 
Wigner function to asymptotic on-shell `out' particle states  \cite{IZ}. In this way, 
we obtain: 
\begin{eqnarray}\label{spectrum} 
2\omega_p\frac{\mbox{d}N}{\mbox{d}^3p}&\equiv&{\cal N}_{(2)}(p)    
=\frac{1}{2(2\pi )^3}\sum_r\;\bar u^r(p)(\gamma\cdot p-m)f_{(2)}(0,p)(\gamma\cdot p-m)u^r(p) 
\\ [1ex]  
&=&\frac{g^2}{4\pi^2}\int\mbox{d}^4q\;\theta (q^0-p^0)\delta \left ((q-p)^2-m^2\right )
\nonumber \\ [1ex] \label{spectrum1}
&\;&\;\;\;\;\;\;\cdot\left (\vec E(q)\cdot\vec E(-q)-\vec B(q)\cdot\vec B(-q)-|A(q)\cdot (q-2p)|^2\right )
\;\;, \end{eqnarray}
where the electromagnetic fields and the vector potential enter; here $\omega_p\equiv\sqrt{\vec p^2+m^2}$   
and $u^r$ ($\bar u^r$) denotes the 
single-particle (adjoint) spinor wave function, respectively, in the normalization of Ref.\,\cite{IZ}. 
Also the last term of this result is gauge invariant, as can be demonstrated using 
the constraints of the integral and the fact that the particles are on-shell. Interestingly, with 
the external fields left completely general, still there is no spatial dependence of this spectrum.
  
Finally, we calculate the \underline{vacuum decay rate} ${\cal R}$  per unit four-volume, 
using a more suitable intermediate form 
of the spectrum result, Eq.\,(\ref{spectrum1}):
\begin{eqnarray}\label{vacR}
{\cal R}&=&2\int\frac{\mbox{d}^3p}{2\omega_p}\;{\cal N}_{(2)}(p)
\\ [1ex] \label{vacR1} 
&=&\frac{g^2}{\pi^2}\int\frac{\mbox{d}^3p}{\omega_p}\frac{\mbox{d}^3p'}{\omega_p'}\mbox{d}^4q\; 
\delta^4(q-p-p')\left (-\frac{1}{2}q^2g_{\mu\nu}+p_\mu p'_\nu+p'_\mu p_\nu\right )
A^\mu (q)A^\nu (-q)
\\ [1ex] \label{vacR2} 
&=&\frac{g^2}{12\pi}\int\mbox{d}^4q\;\theta (q^2-4m^2)\left (1-\frac{4m^2}{q^2}\right )^{1/2}
\left (1+\frac{2m^2}{q^2}\right )\left (|\vec E(q)|^2-|\vec B(q)|^2\right )
\;\;, \end{eqnarray} 
which confirms the result obtained by other methods in Ref.\,\cite{IZ}. Clearly, particle production 
is an electric field effect.   
  
For completeness, let us quote Schwinger's nonperturbative result \cite{Schwinger}:   
\begin{equation}\label{SchwingerR} 
{\cal R}=\frac{g^2\vec E^2}{4\pi^3}\sum_{n=1}^{\infty}\;\frac{1}{n^2}\exp \left ( 
-\frac{n\pi m^2}{|gE|}\right )
\;\;, \end{equation} 
which forms the basis to date of the phenomenological models mentioned in the 
beginning of this section. For reasons which we discussed, a future calculation bridging 
the gap between our perturbative calculation for arbitrary fields and the nonperturbative 
result for constant electric fields would be extremely useful.
 
\section{Fluid dynamics of relativistic quantum dust}
In this section, we will study the relation between relativistic hydrodynamics
and the full quantum evolution of a free matter field \cite{HTE02}. In particular, 
we try to answer how a free fermion field and its energy-momentum tensor 
will evolve, given arbitrary initial conditions and especially those 
of the Landau and Bjorken models.

In the absence of
interactions, decoherence or thermalization may be present in the initial
state, corresponding to an impure density matrix, but is followed by unitary
evolution. We consider this as a ``quantum dust'' model of the expansion of matter
originating from a high energy density preparation phase, which the Landau and Bjorken
models describe classically \cite{Landau,Bj}.

Our approach is independent of the nature of the field, as long as it obeys a
standard wave equation. To be definite, we choose to work with
Dirac fermions and comment about neutrinos later.
We introduce the \underline{spinor Wigner function}, i.e., a (4x4)-matrix depending on
space-time and four-momentum coordinates:
\begin{equation}\label{Wigner}
W_{\alpha\beta}(x;p)\equiv\int\frac{\mbox{d}^4y}{(2\pi )^4}
e^{-ip\cdot y}
\langle :\bar\psi_\beta (x+y/2)\psi_\alpha (x-y/2):\rangle
\;\;, \end{equation}
where the expectation value refers to the (mixed) state of the system;
without interactions, the vacuum plays only a passive role and, therefore,
is eliminated by normal-ordering the field operators. Note that the normalization 
of the Wigner function is a matter of convention, and we have chosen the most 
convenient one for us here.

All observables can be expressed in terms of the Wigner function here.
In particular, the (unsymmetrized) \underline{energy-momentum tensor}:
\begin{equation}\label{Tmunu}
\langle :T_{\mu\nu}(x):\rangle\equiv i\langle :\bar\psi (x)\gamma_\mu
\stackrel{\leftrightarrow}{\partial}_\nu\psi (x):\rangle
=\mbox{tr}\;\gamma_\mu\int\mbox{d}^4p\;p_\nu W(x;p)
\;\;, \end{equation}
where $\stackrel{\leftrightarrow}{\partial}\equiv\frac{1}{2}(\overrightarrow\partial -\overleftarrow\partial )$
and with a trace over spinor indices (conventions as in \cite{VGE87}). Furthermore,
the dynamics of $W$ reduces to the usual phase space description, as in Section 2, in the classical limit \cite{VGE87}.

The following study is based on the simple fact that the 
propagation of the free fields entering in Eq.\,(\ref{Wigner}) from one
time-like hypersurface to another is described by the \underline{Schwinger function}.
It is the solution of the homogeneous Dirac equation,
$[i\gamma\cdot\partial_x-m]S(x,x')=0$, for the initial condition
$S(\vec x,\vec x',x^0=x'^0)=-i\gamma^0\delta^3(\vec x-\vec x')$. Thus,
$\psi (x)=i\int\mbox{d}^3x'S(x,x')\gamma^0\psi (x')$, and similarly for the adjoint.
An explicit form is:
\begin{equation}\label{Schwinger}
iS(x,x')=iS(x-x'\equiv\Delta)=(i\gamma\cdot\partial_{\Delta}+m)
\int\frac{\mbox{d}^3k}{(2\pi )^32\omega_k}\left (
e^{-ik_+\cdot\Delta}-e^{-ik_-\cdot\Delta}\right )
\;\;, \end{equation}
where $k_\pm\equiv (\pm\omega_k,\vec k)$ and $\omega_k\equiv (\vec k^2+m^2)^{1/2}$.

Making use of Eqs.\,(\ref{Wigner}) and (\ref{Schwinger}),
we relate the Wigner function at different times, $t=x^0,x'^0$:
\begin{equation}\label{Wevolution}
W(x;p)=\int\frac{\mbox{d}^4k}{(2\pi )^3}e^{-ik\cdot x}
\delta^\pm (p_k^+)\delta^\pm (p_k^-)
\int\mbox{d}^3x'e^{ik\cdot x'}\int\mbox{d}p'^0
\Lambda(p_k^+)\gamma^0W(x';p')\gamma^0\Lambda(p_k^-)
\;\;, \end{equation}
where $p_k^\pm\equiv p\pm\frac{k}{2}$, $\delta^\pm (q)\equiv\pm\delta(q^2-m^2)$ (for $q^0/|q^0|=\pm 1$),
$\Lambda (q)\equiv\gamma\cdot q+m$, and $p'^\mu\equiv (p'^0,\vec p)$.

The Eq.\,(\ref{Wevolution}) implies that the Wigner function obeys a generalized \underline{mass-shell
constraint} and a proper \underline{free-streaming transport equation}:
\begin{eqnarray}\label{constraint}
[p^2-m^2-\frac{\hbar^2}{4}\partial_x^2]\;W(x;p)&=&0 \;\;, \\ [2ex]
\label{transport} p\cdot\partial_x\;W(x;p)&=&0 \;\;,
\end{eqnarray} separately for each matrix element. The reinserted
$\hbar$ indicates the important quantum term in the equations,
which otherwise have the familiar classical appearance.

Thus Eq.\,(\ref{Wevolution}) presents an \underline{integral solution} of the microscopic transport equations
for a given initial Wigner function. Furthermore, a semiclassical approximation of the Schwinger function
may be used to generate an integral solution of the corresponding classical transport problem.

Next, we decompose the Wigner function with respect to the standard basis of
the \underline{Clifford algebra}, $W={\cal F}+i\gamma^5{\cal P}+\gamma^\mu {\cal V}_\mu
+\gamma^\mu\gamma^5{\cal A}_\mu +\frac{1}{2}\sigma^{\mu\nu}{\cal S}_{\mu\nu}$,
i.e., in terms of scalar, pseudoscalar, vector, axial vector, and
antisymmetric tensor components \cite{IZ}. The functions,
${\cal F}\equiv\frac{1}{4}\mbox{tr}\;W$\,,
${\cal P}\equiv-\frac{1}{4}i\;\mbox{tr}\;\gamma^5 W$\,,
${\cal V}_\mu\equiv\frac{1}{4}\mbox{tr}\;\gamma_\mu W$\,,
${\cal A}_\mu\equiv\frac{1}{4}\mbox{tr}\;\gamma_5\gamma_\mu W$\,,
and ${\cal S}_{\mu\nu}\equiv\frac{1}{4}\mbox{tr}\;\sigma_{\mu\nu}W$\,,
which represent physical current densities,
are real, due to $W^\dagger =\gamma^0W\gamma^0$ \cite{VGE87}. They individually obey
Eqs.\,(\ref{constraint}) and (\ref{transport}).

We assume ${\cal P}=0={\cal A}_\mu$, i.e., we consider a spin saturated system for simplicity. 
-- From this point on, a corresponding study of (approximately) massless Standard Model
$\nu_L\bar\nu_R$ neutrinos differs, and is simpler, since ${\cal V}_\mu ={\cal A}_\mu$\,,
while all other densities vanish identically; see, e.g., Ref.\,\cite{neutrino}. -- 
Then, using the `transport equation' which follows directly from the
Dirac equation applied to $W$, $[\gamma\cdot (p+\frac{i}{2}\partial_x)-m]W(x;p)=0$, and
decomposing it accordingly, the following additional relations among the remaining densities
are obtained:
\begin{eqnarray}\label{vector}
{\cal V}^\mu (x;p)&=&\frac{mp^\mu}{p^2}{\cal F}(x;p)
\;\;, \\ [2ex]
\label{tensor}
{\cal S}^{\mu\nu}(x;p)&=&\frac{1}{2p^2}(p^\nu\partial_x^\mu -p^\mu\partial_x^\nu ){\cal F}(x;p)
\;\;. \end{eqnarray}
Note that ${\cal S}^{\mu\nu}$ is intrinsically by one order in $\hbar$ smaller than the
other two densities.

We conclude that presently the dynamics of the system is represented completely by the \underline{scalar} phase space
\underline{density} ${\cal F}$, which obeys the same transport equations as the full $W$ itself. 
Using Eqs.\,(\ref{Tmunu}) and (\ref{vector}), we obtain in particular:
\begin{equation}\label{TF}
\langle :T^{\mu\nu}(x):\rangle =4m\int\mbox{d}^4p\;\frac{p^\mu p^\nu}{p^2}{\cal F}(x;p)
\;\;, \end{equation}
which is symmetric and conserved, $\partial_\mu T^{\mu\nu}(x)=0$, on account of Eq.\,(\ref{transport}).
Furthermore, this implies the `\underline{equation of state}':  
\begin{equation}\label{eos}
\langle :T^{00}(x):\rangle -\sum_{i=1}^3\langle :T^{ii}:\rangle 
=4m\int\mbox{d}^4p\;{\cal F}(x;p)
\;\;, \end{equation}
which relates energy density and pressure(s). However, applying Eq.\,(\ref{constraint}), we find that this relationship evolves in a wavelike manner, driven by off-shell contributions to the evolving ${\cal F}$:
\begin{equation}\label{eoswave}
\partial_x^2\langle :T_\mu^\mu (x):\rangle =16m\int\mbox{d}^4p\;
(p^2-m^2){\cal F}(x;p)
\;\;. \end{equation}
This differs from classical hydrodynamics with a fixed functional form of the equation of state.  
Eqs.\,(\ref{TF})-(\ref{eoswave}) hold independently of the initial state, of course, if  
it evolves without further interaction.  

Making use of  Eq.\,(\ref{Wevolution}) in Eq.\,(\ref{TF}), we now calculate the
energy-momentum tensor at any time in terms of the initial scalar density. Employing the decomposition
of the Wigner function and commutation and trace relations for the $\gamma$ matrices, as well as
Eqs.\,(\ref{constraint})-(\ref{tensor}), we obtain:
\begin{eqnarray}
&\;&\langle :T^{\mu\nu}(x):\rangle =8m\int\dots\int
\frac{p^\mu p^\nu}{p^2}
\left (p^2+\frac{m^2}{p'^2}\left (p^0p'^0+\vec p^{\;2}\right )
-\frac{1}{4p'^2}\left ((p^0k^0)^2-\vec p^{\;2}\vec k^2 \right .
\right . \nonumber
\\[1ex]\label{TmunuEvolution}
&\;&\left . \;\;\;\;\;\;\;\;\;\;\;\;\;\;\;\;\;\;\;\;\;\;\;\;\;\;\;\;
\;\;\;\;\;\;\;\;\;\;\;\;\;\;\;\;\;\;\;\;\;\;\;\;
+p^0p'^0k^2+\frac{p^0}{p'^0}\left .(k^0)^2p^2\right )
\right ){\cal F}(x';\vec p,p'^0)
\;\;, \end{eqnarray}
where $\int\dots\int\equiv (2\pi )^{-3}
\int\mbox{d}^4p\int\mbox{d}^4k\;e^{-ik\cdot x}
\delta^\pm (p_k^+)\delta^\pm (p_k^-)
\int\mbox{d}^3x'e^{ik\cdot x'}\int\mbox{d}p'^0$; we also made use of partial integrations and
the $\delta$-function constraints. The three terms on the right-hand side stem from
the scalar, vector, and antisymmetric tensor
components of the initial Wigner function, respectively. 
 
If the initial distribution is an isotropic function of the three-momentum,
then $T^{\mu\nu}$ is diagonal at all times, implying that the absence of flow 
in the initial state will be preserved.

Indeed, we expect the \underline{(non-)flow features} of the initial
distribution to be preserved during the evolution, due to the
absence of interactions. Kinetic energy from microscopic
particle degrees of freedom will not be converted into collective
motion. An interesting question is, how the classical hydrodynamic
acceleration of fluid cells due to pressure gradients arises in
our present model after coarse graining
\cite{WHZ90,I95,Lisewski99,BrunHartle99}. This has not been studied yet. 
Recalling earlier work on the hydrodynamic 
representation of quantum mechanics, e.g. Refs.\,\cite{Madelung},
and recently deduced classical fluid behavior of quantum fields in
WKB approximation \cite{DomLev00}, however, we study the full 
quantum evolution here.

We are particularly interested in the exact evolution of $T^{\mu\nu}$,   
assuming a particle-antiparticle symmetric initial state. This is believed to hold,
for example, close to midrapidity in the center-of-mass frame of central high-energy collisions \cite{Bj,QMs}.
It implies that the initial ${\cal F}$ is an even function of
the energy variable, ${\cal F}(x';\vec p,p'^0)={\cal F}(x';\vec p,-p'^0)$.  
While Eq.\,(\ref{TmunuEvolution}) allows general initial conditions, we follow 
the implicit \underline{on-shell} assumption in classical hydrodynamic models:
\begin{equation}\label{onshell}
{\cal F}(x';\vec p,p'^0)=(2\pi )^{-3}m\;\delta (p'^2-m^2)\left (
\Theta (p'^0)F(x';\vec p,p'^0)+\Theta (-p'^0)F(x';\vec p,-p'^0)\right )
\;\;. \end{equation}
Fermion \underline{blackbody radiation} is described by
$F(x';\vec p,p'^0)\equiv f(p'^0/T(x'))$,
where $T$ denotes the local temperature, and with $f(s)\equiv (e^s+1)^{-1}$; this is easily
illustrated with the help of Eqs.\,(\ref{TF}) and (\ref{onshell}). 

Implementing Eq.\,(\ref{onshell}), we obtain the simpler result:
\begin{eqnarray}\label{Tmunuonshell}
&\;&\langle :T^{\mu\nu}(x):\rangle =\int\frac{\mbox{d}^3x'\mbox{d}^3p}{(2\pi )^3}\frac{\mbox{d}^3k}{(2\pi )^3}
\;\frac{p^\mu p^\nu\cos [\vec k\cdot (\vec x-\vec x')]}{\omega_p\omega_+\omega_-}
F(x';\vec p,\omega_p)
\\ [1ex]
&\;&\cdot\{ ((\omega_++\omega_-)^2-\vec k^2 )\cos [(\omega_+-\omega_-)t]
-((\omega_+-\omega_-)^2-\vec k^2)\cos [(\omega_++\omega_-)t] \}
\;\;, \nonumber \end{eqnarray}
where $t\equiv x^0-x'^0$, $\omega_p\equiv (\vec p^2+m^2)^{1/2}$, $\omega_\pm\equiv ((\vec p\pm\vec k/2)^2+m^2)^{1/2}$;  
furthermore, $p^0\equiv\frac{1}{2}|\omega_+\pm\omega_-|$, with ``+'' when multiplying the first and 
``$-$'' when multiplying the second term of the difference, respectively. 
Depending on geometry and initial state, further
integrations can be done analytically.

Consider a \underline{$(1+1)$-dimensional system} for illustration, 
assuming that the particles are approximately massless, i.e. $\omega_p\approx |p|$, and that 
$F$ is even in $p$ (no flow).   
Specializing to a kind of \underline{Landau initial condition}, the distribution is 
prepared on a fixed timelike hypersurface at $t=0$ \cite{Landau}. We    
find the ultrarelativistic equation of state for the only 
nonvanishing components of $T^{\mu\nu}$, $\epsilon\equiv T^{00}=T^{11}\equiv P$ ($d=1+1$), 
which are calculated as a momentum integral following Eq.\,(\ref{Tmunuonshell}): 
\begin{eqnarray}\label{Landau} 
T^{00}(x,t)&=&2\int\frac{\mbox{d}p}{2\pi}|p|{\displaystyle\Bigl (
F(x-t;|p|)+F(x+t;|p|)\Bigr )}
\\ [1ex] \label{LandauWave} 
&=&\frac{1}{2}\left (T^{00}(x-t,t=0)+T^{00}(x+t,t=0)\right ) 
\;\;, \end{eqnarray} 
i.e., a superposition of wavelike propagating momentum contributions 
in accordance with Eq.\,(\ref{eoswave}). 

Similarly, a kind of \underline{Bjorken initial condition} can be specified on a surface of 
constant proper time \cite{Bj}. A transformation of Eq.\,(\ref{Landau}) 
to space-time rapidity and proper time coordinates yields:  
\begin{equation}\label{Bjorken} 
T^{00}(y,\tau )=2\int\frac{\mbox{d}p}{2\pi}|p|{\displaystyle\Bigl (
F(-\tau_0 e^{-y/2+\ln \tau /\tau_0};|p|)+F(\tau_0 e^{y/2+\ln \tau /\tau_0};|p|)\Bigr )} 
\;\;, \end{equation}
since $x\equiv\tau\sinh y/2$ and $t\equiv\tau\cosh y/2$ ($\tau\geq \tau_0>0$). 

Our results for free fermions show the free-streaming behavior of \underline{classical dust}, 
associated with the independent propagation and linear superposition of the momentum  
contributions to the scalar component $F$ of the Wigner function here. In particular, 
the shape function of each mode is preserved and translated lightlike  
(with dispersion for massive particles). 
Due to the assumed momentum symmetry, the initial distribution will separate into two 
components after a finite time, travelling into the forward and 
backward direction, respectively, with a corresponding dilution at the center. For example, 
an initial distribution of Gaussian shape will separate into two corresponding humps.  
    
We recall that $T^{\mu\nu}$ being diagonal implies 
the absence of ideal hydrodynamic flow, given $\epsilon =(d-1)P$. This does not depend 
on whether the initial state is on- or off-shell, see Eq.(\ref{TmunuEvolution}).   
Therefore, any hydrodynamic 
behavior must be the effect of a pecularity of the semiclassical limit \cite{DomLev00}, 
of coarse graining \cite{I95,Lisewski99,BrunHartle99},  
or of interactions \cite{Cooper01}, or a combination of these.  
   
Despite the apparently classical evolution, however, all initial state  
\underline{quantum effects} are incorporated and preserved. If the 
initial dimensionless distribution $F$ has a dependence on products of 
momentum and space-time variables, which is characteristic for matter waves, such terms 
invoke a factor $1/\hbar$. Similarly, if it is thermal ($T$) but includes the 
finite size ($L$) shell effects or global constraints, then there are quantum 
corrections involving $LT/\hbar$ ($k_B=c=1$) \cite{IG}. For typical values of 
$LT/\hbar\approx 1$ these latter corrections are known to lead to corrections on the order 
of 30\% in the thermodynamical quantities. They have not been 
included in semiclassical transport or classical hydrodynamic models of high-energy 
(nuclear) collisions, but may be large. Here the quantum dust model provides  
a valuable testing ground to assess the importance of these quantum effects. 
  
Let us summarize briefly the results and perspective of this section: 
\begin{itemize}
\item Based on the Schwinger function, we obtain the solution of the 
free quantum transport problem to quadratures for arbitrary on- or off-shell initial conditions. 
\item In 1+1 dimensions free fermions, i.e. their observables embodied in the 
energy-momentum tensor, collectively behave like classical dust showing no flow effects, if 
there is no flow present in the initial condition.  
Corresponding analytical results in 
three dimensions can be obtained and will be discussed elsewhere. 
\item The method presented here may lead 
to an efficient way of treating interacting particles. Especially when a low-order 
perturbative expansion is meaningful, interactions could be incorporated in a multiple 
scattering expansion. With free propagation in 
between scattering events, a consistent and conceptually simple space-time description of 
transport phenomena seems feasible.
\end{itemize}   
  
Thus we conclude our introductory lectures on relativistic quantum transport theory, 
which should convey some of its basic concepts and hopefully will lead to some of the 
interesting topics for further study.
     
\subsection*{Acknowledgement} 
I wish to thank A.G.\,Grunfeld and A.M.S.\, Santo for a careful reading and 
their helpful suggestions improving the manuscript.  


\end{document}